\documentclass[12pt]{article}

\usepackage{amsmath,amsfonts, amssymb}

\begin{document}
\bibliographystyle{abbrv}

\title{The von Neumann entropy and information rate 
for integrable quantum Gibbs ensembles, 2}
\author{Oliver Johnson\thanks{Statistical Laboratory, 
DPMMS/CMS, University of Cambridge, Cambridge CB3 0WB, UK. 
Fax: +44 1223 337956} \thanks{Email:\tt{otj1000@cam.ac.uk}} 
\and Yuri Suhov$^*$\thanks{Email:\tt{yms@statslab.cam.ac.uk}}}
\date{\today}
\maketitle

\newtheorem{theorem}{Theorem}[section]
\newtheorem{lemma}[theorem]{Lemma}
\newtheorem{proposition}[theorem]{Proposition}
\newtheorem{corollary}[theorem]{Corollary}
\newtheorem{conjecture}[theorem]{Conjecture}
\newtheorem{definition}[theorem]{Definition}
\newtheorem{example}[theorem]{Example}
\newtheorem{condition}{Condition}
\newtheorem{main}{Theorem}
\newtheorem{assumption}[theorem]{Assumption}
\renewcommand{\theequation}{\arabic{section}.\arabic{equation}}
\setlength{\parskip}{\parsep}
\setlength{\parindent}{0pt}

\def \outlineby #1#2#3{\vbox{\hrule\hbox{\vrule\kern #1%
\vbox{\kern #2 #3\kern #2}\kern #1\vrule}\hrule}}%
\def \endbox {\outlineby{4pt}{4pt}{}}%
\newenvironment{proof}
{\noindent{\bf Proof\ }}{{\hfill \endbox
}\par\vskip2\parsep}
\newenvironment{pfof}[2]{\removelastskip\vspace{6pt}\noindent
 {\it Proof  #1.}~\rm#2}{\par\vspace{6pt}}

\hfuzz12pt

\newcommand{\Section}[1]{\setcounter{equation}{0} \section{#1}}
\newcommand{\cl}[1]{{\mathcal{C}}_{#1}}
\newcommand{\var}{{\rm{Var\;}}}
\newcommand{\cov}{{\rm{Cov\;}}}
\newcommand{\tends}{\rightarrow \infty}
\newcommand{\ep}{{\mathbb {E}}}
\newcommand{\pr}{{\mathbb {P}}}
\newcommand{\re}{{\mathbb {R}}}
\newcommand{\I}{\mathbb {I}}
\newcommand{\vc}[1]{{\bf {#1}}}
\renewcommand{\vec}[1]{\vc{#1}}
\newcommand{\ra}[2]{{#1}^{(#2)}}
\newcommand{\Z}{{\mathbb {Z}}}
\newcommand{\R}{{\mathbb {R}}}
\newcommand{\K}{{\mathbb {K}}}
\newcommand{\ent}{{\rm E}}
\newcommand{\bin}[2]{\binom{#1}{#2}}
\newcommand{\ov}[1]{\overline{#1}}

\begin{abstract}
This paper considers the problem of data compression for dependent
quantum systems. It is the second in  a series under the same title which began
with   \cite{johnson10}  and   continues  with   \cite{ssy}.    As  in
\cite{johnson10}, we  are interested  in  Lempel--Ziv  encoding for
quantum  Gibbs  ensembles.   Here,  we consider  the  canonical  ideal
lattice Bose- and Fermi-ensembles. We prove that as in the case of the
grand canonical ensemble, the  (limiting) von Neumann entropy rate $h$
can  be  assessed,  via  the classical  Lempel--Ziv  universal  coding
algorithm,  from a  single eigenvector  $\psi$ of  the  density matrix
$\rho$.
\end{abstract}

\section{Introduction}\label{sec:introduction}

This paper continues paper \cite{johnson10} under the same title and
extends results established there for
grand canonical ensembles to canonical ensembles of
ideal (free) quantum systems, bosonic or fermionic.
This will allow us to analyse the question of 
the Bose-Einstein condensation (in the case of a bosonic
ensemble) and extend results to a number of
other integrable models (see \cite{ssy}).

The reader is referred to \cite{johnson10} for a general introduction
into the subject; here we only provide a background 
formally needed for exposition of the results. 

We consider ideal quantum systems on a cubic lattice $\Z^d$.
The starting point here is a given function $\omega$:
${\mathbf y}\in[0,1]^d\mapsto\omega (y)$ (more precisely, a family of functions
$\omega_\mu$ depending on the chemical potential $\mu$), with
non-negative values for bosons and real for fermions, 
describing the energy of normal mode ${\mathbf y}$ 
(in a single-particle momentum space $[0,1]^d$).
An example which we will follow closely is where
\begin{equation}
\omega_{\mu} (\vec{y}) 
=\frac{1}{d}\sum\limits_{1\leq l\leq d}[1-
\cos\,(2\pi y_j)]-\mu ,\;\;
\vec{y}=(y_1,\ldots ,y_d)\in[0,1]^d,\label{eq:sampom} \end{equation} 
and $\mu <0$ for bosons, $\mu\in\R$ for fermions.
Function (\ref{eq:sampom}) determines the 
Fourier transform of the operator 
$-\frac{1}{2}\Delta -\mu I$ 
acting in $l_2(\Z^d)$ where $\Delta$ stands for 
the discrete Laplacian on $\Z^d$.

Associated with $\omega_{\mu} ({\mathbf y})$ are two 
important integrals:
\begin{equation}
h^\pm_{\beta, \mu} =\int_{[0,1]^d} \left( \mp \log
\left(1 \mp e^{-\beta \omega_{\mu}(\vec{y})} \right) + 
\frac{\beta \omega_\mu(\vec{y})}{\log e}  
\frac{ e^{-\beta \omega_\mu(\vec{y})}}{1 \mp e^{-\beta \omega_\mu(\vec{y})}} 
\right)  d\vec{y}\label{eq:entdef} \end{equation}
and 
\begin{equation}
m^\pm_{\beta,\mu} = \int_{[0,1]^d}
\frac{ e^{-\beta \omega_\mu(\vec{y})}}{1 \mp e^{-\beta \omega_\mu(\vec{y})}} 
d\vec{y}\label{eq:partden} \end{equation}
giving, 
respectively, the value of the von Neumann entropy rate
and the particle density, in the thermodynamic limit. 
(In the left-hand side we put plus for bosons 
and minus for fermions; we will follow this convention throughout
the paper). Parameter $\beta >0$ is the inverse temperature.

To simplify matters, we will assume that $d=1$, 
though all technicalities can be extended to the
case of a general $d$ in a straightforward way.\footnote{
The issue of Bose-Einstein condensation arises 
of course for $d\geq 3$, unless one uses particular boundary 
conditions. This will be the subject of forthcoming research.}
The free Gibbs grand-canonical Bose- or Fermi-ensemble 
in a finite volume $\Lambda_\ell=$ 
$\{0,1,\ldots, \ell -1\}$ (represented by a
segment of the integer lattice $\Z$) is described
by a density matrix $\rho_\ell^\pm$ acting in the Fock Hilbert space
${\mathcal H}_\ell^\pm$. It is of the form
$$\rho_\ell^\pm=\frac{1}{\Xi^\pm_\ell}\exp (-\beta H^\pm_\ell)$$
where $\Xi^\pm_\ell={\rm{tr}}_{{\mathcal H}^\pm_\ell}
\exp (-\beta H^\pm_\ell )$
and $H^\pm_\ell$ is the Hamiltonian of the free
Bose- or Fermi-system in $\Lambda_\ell$ which
is the (bosonic or fermionic) second quantisation 
of the single-particle
energy operator $H_{\ell, 1}$ in $l_2(\Lambda_\ell )
\simeq {\mathbb C}^\ell$. In the above example,
\begin{equation}
H_{\ell, 1}= -\frac{1}{2}\Delta_\ell-\mu I_\ell \end{equation}
where $\Delta_\ell$ is the lattice Laplacian in
$\Lambda_\ell$  and $I_\ell$ the unit matrix in $l_2(\Lambda_\ell )$. If 
we impose periodic boundary conditions then
the eigenvectors $\psi_\ell$ and eigenvalues $\kappa_\ell$
of $H_{\ell, 1}$ are naturally labelled by $j=0,1,\ldots, \ell -1$:
\begin{equation}
\varphi_\ell(j)=\frac{1}{\sqrt{\ell}}\exp (2\pi ij/\ell),
\;\;\kappa_\ell (j)=1-\cos (2\pi ij/\ell )-\mu.\label{eq:eigenval}
\end{equation}
In particular, the eigenvalues $\kappa_\ell$ in (\ref{eq:eigenval}) 
are described as $\omega_\mu (j/\ell )$ where function $\omega$
was defined in (\ref{eq:sampom}).

We retain the notation $\psi_\ell(j)$ and $\kappa_\ell(j)$ for the eigenvectors
and eigenvalues of a genereal single-particle energy operator $H_{\ell,1}$ in
$l_2(\Lambda_\ell)$, and assume the form $\kappa_\ell(j) = \omega_\mu(j/\ell)$
($j = 0, 1, \ldots \ell -1$), where $\omega_\mu(y)$, is a `nice' function
figuring in (\ref{eq:sampom}) and (\ref{eq:entdef}). See Assumption 
\ref{ass:mu} below.

The condition that $\kappa_\ell(j)$ is of the form $\omega_\mu(j/\ell)$ is
quite restrictive (although it holds in the example of (\ref{eq:partden})
with periodic boundary conditions). We consider it as a first step in studying
more general situations.

Returning to the multi-particle Hamiltonian $H_{\ell}^{\pm}$, its eigenvectors
$\phi_{\ell}$ and eigenvalues $\lambda_\ell$ are naturally labelled by
occupancy number configurations $\vc{k} = (k_0, \ldots k_{\ell-1})$
where entry $k_j\in\{0,1\}$ for fermions and
$k_j\in\Z_+$ for bosons:
$$\vc{k}\in\{0,1\}^\ell\;\hbox{ or }\;
\vc{k}\in\Z^\ell .$$ 
Space ${\mathcal H}_\ell^-$ has dimension
$2^\ell$ and ${\mathcal H}_\ell^+$ infinite dimension.

More precisely, the eigenvectors and eigenvalues of $H_{\ell}^{\pm}$ have 
the form
\begin{equation}
\phi^\pm_\ell (\vc{k})=\left(\prod\limits^\otimes_{j\in
\Lambda_\ell}\psi_\ell (j)^{\otimes k_j}\right)_{S/A},\;\;\;
\lambda^\pm_\ell =\sum\limits_{j\in\Lambda_\ell}k_j\kappa_\ell (j).
\label{eq:occnum} \end{equation}
Here, subscript $S$ means symmetrisation and $A$
antisymmetrisation of the corresponding tensor
product $\prod\limits^\otimes_{j\in
\Lambda_\ell}\psi_\ell (j)^{\otimes k_j}$.

In probabilistic terms, the density matrix
$\rho^\pm_\ell$ generates a probability distribution 
where an eigenvector $\phi_\ell(\vc{k})$ `occurs'
with probability $\frac{1}{\Xi^\pm_\ell}\exp\;(-\beta\lambda_\ell 
(\vc{k}))$, with $\Xi^\pm_\ell=$ 
$\sum\limits_{{\widetilde{\vc{k}}}\in\Lambda_\ell}\exp\;(
-\beta\lambda_\ell({\widetilde{\vc{k}}}))$. It is convenient
to assign the above probability to occupancy number
configuration $\vc{k}$ i.e. consider a probability 
measure ${\mathcal P}^\pm_\ell$ on space $\Z^\ell$ (bosons) or
$\{0,1\}^\ell$ (fermions) 
$${\mathcal P}^\pm_\ell (\vc{k})=\frac{1}{\Xi^\pm_\ell}
\exp\;(-\beta\lambda_\ell(\vc{k})).$$
As follows from Equation (\ref{eq:occnum}), the entries $K_j$,
$j\in\Lambda_\ell$, of the random configuration 
$\vc{K}=(K_0,\ldots ,K_{\ell -1})$
are independent variables, each with a two-point 
distribution for fermions and geometric for bosons:       
\begin{equation}
{\mathcal P}^\pm_\ell (K_j=k)=(1\mp e^{-\beta\kappa_\ell (j)})^{\pm 1}
e^{-k\beta\kappa_\ell (j)},\;\;\begin{array}{l}
k \in \Z_+ ,\\ k=0,1.\end{array}\label{eq:margdef} \end{equation}

The above independence property is a feature of the grand 
canonical ensemble for free particles. 
Paper \cite{johnson10} focused on properties of the Lempel-Ziv encoding for
(the sequence of) probability measures ${\mathcal P}_\ell$
as $\ell\to\infty$ (the thermodynamic limit).
The main result of \cite{johnson10} was that if (i) eigenvalues
$\kappa_\ell (j)$ of the single-particle energy
operator $H_{\ell, 1}$ are of the form $\omega (j/\ell)$
(which is the case in the example under consideration)
and (ii) function 
$\omega$ satisfies a certain condition (see Assumption
1 from Section 2 of \cite{johnson10}), the Lempel-Ziv parsing of the
random string $\vc{K}=(K_0,\ldots ,K_{\ell -1})$
indentifies integral (\ref{eq:entdef}) as the data compression limit.

Replacing the grand canonical with the canonical
ensemble means fixing values $n_\ell$, $\ell
=1,2$, ... for the number of particles. Probabilistically,
we have to pass to conditional distributions
${\mathcal P}^{\pm}_{\ell,n_{\ell}} = 
{\mathcal P}^\pm_\ell\left(\;\cdot\;\left|\sum\limits_{0 \leq j \leq n-1} 
K_j=n_\ell\right.\right)$ which do not
have the independence property.    

It is convenient to specify the integrands
in (\ref{eq:entdef}) and (\ref{eq:partden}) as
\begin{equation}
g^\pm_{\beta,\mu}(y) =
-\left(\pm \log
\left(1 \mp e^{-\beta \omega_\mu (y)} \right)\right)+ 
\frac{\beta \omega_\mu (y)}{\log e}  
\frac{ e^{-\beta \omega_\mu (y)}}{1 \mp e^{-\beta 
\omega_\mu (y)}},\; 0 \leq y \leq 1,\label{eq:entdef2} \end{equation}
and 
\begin{equation}
l^\pm_{\beta,\mu}(y) =
\frac{e^{-\beta \omega_\mu (y)}}{1\mp e^{-\beta \omega_\mu (y)}},
\;0 \leq y \leq 1,\label{eq:partden2} \end{equation}
and interpret the values $l_{\beta,\mu}^{\pm}(j/\ell )$ and 
$g^\pm_{\beta,\mu}(j/\ell )$  
as the mean and entropy of the marginal probability distribution
(two-point or geometric) 
of variable $K_j$ in (\ref{eq:margdef}), $j\in\Lambda_\ell$. [Of course,
these values can be expressed in terms of each other.] 
Here, in agreement with (\ref{eq:sampom}), we set:
$$\omega_\mu (y)=\omega_0(y)-\mu,$$
although a more general form of dependence can be considered.
We also assume that $\omega_0$ is a non-negative function 
for bosons and real for fermions. 

Observe that while parameter $\mu$ figures in the definition of the 
measure ${\mathcal P}_\ell^{\pm}$, it does not in that of 
the conditioned measure ${\mathcal P}_{\ell,n_{\ell}}^{\pm}$.

From now on, we state conditions that we need in terms of 
the mean-value functions 
$l^\pm_{\beta,\mu}(y)$, and the whole exposition is purely probabilistic.    
We follow the notation and definitions fron Sections 1 and 
2 of \cite{johnson10}.

\begin{assumption} \label{ass:mu}
Functions $l^\pm_{\beta,\mu}$ in (\ref{eq:partden2})
take
$$\begin{array}{l}y\in [0,1]\mapsto l^+_{\beta, \mu}(y)\in 
(0,\infty )\\
y\in [0,1]\mapsto l^-_{\beta, \mu}(y)\in (0,1) 
\end{array}$$ 
are continuous in $y$
and for all $\beta >0$ and $\mu <0 $ for bosons and
$\mu\in\R$ for fermions. 
Moreover, for all $\beta >0$, the function
$\mu\mapsto m^{\pm}_{\beta,\mu}$ is monotone increasing
with $\mu$ so that, for all $r$ in the range of this
function, there exists a unique $\mu$ 
such that $m^{\pm}_{\beta,\mu}=r$. Here, and below
\begin{equation}
m_{\beta,\mu}^{\pm} =\int_{[0,1]} l^{\pm}_{\beta,\mu}(y) dy
.\end{equation}
\end{assumption}
Furthermore, functions $$y \in [0,1] \mapsto g_{\beta,\mu}^{\pm}(y) \in
(0, \infty)$$
are continuous in $y$ for all $\beta > 0$ and $\mu < 0$ for bosons and 
$\mu \in \re$ for fermions.

\begin{assumption} \label{ass:nogrow}
Sequence of particle numbers
$n_\ell$ satisfies:
\begin{equation}\left|\frac{n_\ell}{\ell}-r\right| =
o\left(\frac{1}{\sqrt\ell}\right) \end{equation} 
for some $r$ from the range of function 
$\mu\mapsto m^{\pm}_{\beta ,\mu}$. \end{assumption}
Assumptions \ref{ass:mu} and \ref{ass:nogrow} are henceforth presumed to hold.
The main result of the paper is the following:
\begin{theorem} \label{thm:main}
Consider a triangular array of independent random variables 
$K_j^{(\ell)}$, $0 \leq i \leq \ell-1$, $\ell = 2,3, \ldots$,  
(either geometric or $0,1$-valued), where $K_j^{(\ell)}$
has mean $l_{\beta,\mu}^{\pm}(j/\ell)$ and entropy 
$g_{\beta,\mu}^{\pm}(j/\ell)$.
From $K_j^{(\ell)}$, define random variables $Y_j^{(\ell)}$,
distributed as $K_j^{(\ell)} \Big| \left( \sum_{0 \leq i \leq \ell -1} 
K_i^{(\ell)} = n_{\ell} \right)$.

Then for all $\beta > 0$,
in probability, almost surely and in mean the number of words 
$C(\vec{Y}^{(\ell)})$ in the 
Lempel--Ziv parsing of the string $\vec{Y}^{(\ell)} = \{ Y_1^{(\ell)}, \ldots
Y_{\ell}^{(\ell)} \}$ satisfies:
\begin{equation} \label{eq:main}
 \lim_{\ell \tends} \frac{\log \ell}{\ell} C({\vec{Y}}^{(\ell)}) =
h^{\pm}_{\beta,\mu} := \int_{[0,1]} g_{\beta,\mu}^{\pm}(y) dy,
\end{equation}
where
$\mu$ is the unique value for which $m^{\pm}_{\beta ,\mu} =r$. Here, almost
surely is understood with respect to the product-measure $\times_\ell
{\cal P}^{\pm}_{\ell,n_{\ell}}$ (see \cite{johnson10}).
\end{theorem}

The almost sure form of convergence is the most subtle, so we treat it as 
principal. The logic behind the proof is as follows.
By Shannon's Noiseless
Coding Theorem (see for example Theorem 5.3.1 of \cite{cover}), the 
limit of the Shannon entropy of probability distributions 
${\cal P}_{\ell,n_{\ell}}^{\pm}$
$$ \lim_{\ell \tends} \frac{1}{\ell} H \left( K_0, \ldots, K_{\ell-1}
\left| \sum_{0 \leq i \leq \ell-1} K_i = n_{\ell} \right) \right.$$
(if it exists) represents the data compression limit for the sequence of
canonical ensemble distributions ${\cal P}_{\ell,n_{\ell}}^{\pm}$. In terms
of the Lempel-Ziv encoding, repeating arguments from Chapter II of
\cite{shields3} yields:
\begin{lemma} 
Almost surely with respect to $\times {\cal P}_{\ell,n_{\ell}}^{\pm}$:
\begin{equation}
\liminf_{\ell \tends} \frac{\log \ell}{\ell} C( \vec{Y}^{(\ell)})
\geq \liminf_{\ell \tends} \frac{1}{\ell} H \left( K_0, \ldots, K_{\ell-1}
\left| \sum_{0 \leq i \leq \ell-1} K_i = n_{\ell} \right) \right.. 
\label{eq:bdfrombel}
\end{equation}
\end{lemma}

On the other hand, we establish two lemmas:
\begin{lemma} \label{lem:firstbd}
Almost surely with respect to $\times {\cal P}_{\ell,n_{\ell}}^{\pm}$:
\begin{equation}
\limsup_{\ell \tends} \frac{\log \ell}{\ell} C( \vec{Y}^{(\ell)})
\leq h_{\beta,\mu}^{\pm} = 
\lim_{\ell \tends} \frac{1}{\ell} H \left( K_0, \ldots, K_{\ell-1} \right),
\end{equation}
the averaged Shannon entropy of probability measure ${\cal P}_{\ell}$, where
$\mu$ has been specified in Theorem \ref{thm:main}.
\end{lemma}
\begin{lemma} \label{lem:secondbd}
The entropies obey the bound:
\begin{equation}
H \left( K_0, \ldots, K_{\ell-1}
\left| \sum_{0 \leq i \leq \ell-1} K_i = n_{\ell} \right) \right. - 
H \left( K_0, \ldots, K_{\ell-1} \right) \geq \delta(\ell, n_{\ell}), 
\end{equation}
where
\begin{equation}
\lim_{\ell \tends} \frac{1}{\ell} \delta(\ell, n_{\ell}) = 0,
\label{eq:deltazero} 
\end{equation} \end{lemma}
Together (\ref{eq:bdfrombel})--(\ref{eq:deltazero}) imply (\ref{eq:main}).
In Sections \ref{sec:typset} and \ref{sec:na} we develop an argument that
proves Lemma \ref{lem:firstbd} and in Section \ref{sec:vnaens} we prove
Lemma \ref{lem:secondbd}. To simplify the notation, we will often omit
superscripts $\pm$ and subscripts $\beta,\mu$.
\Section{Properties of the typical set} \label{sec:typset}
Throughout the rest of the paper, Assumptions \ref{ass:mu} and 
\ref{ass:nogrow} are presumed valid. In this section 
we concentrate on the fermionic case of $0,1$-
valued variables -- the proofs adapt to the geometric case as in 
\cite{johnson10}.
As in \cite{johnson10}, we will use the idea of a typical set
-- firstly in Proposition \ref{prop:intyp}
we shall show that the added restriction of being in the typical set forces
extra useful properties to hold. Then in Proposition 
\ref{prop:outtyp}, we shall
show that we will `nearly always' be in the typical set, so we can exploit
these extra properties.

We write $e_a$ for the entropy of the random variable under consideration
(geometric or $0,1$-valued) with mean $a$.
Given $\epsilon\in (0,1)$ and 
integer $M$ and $j$, 
define the typical set of realisations ${\mathcal{T}}^{(\ell)}_{j,M}$ 
by 
\begin{equation} \label{eq:typset} {\mathcal{T}}^{(\ell)}_{j,M}= 
\left\{ \vec{k}: 
\sum_{i=j}^{j+M-1} (k_i- l(i/\ell)) \leq M \epsilon' \right\},
\mbox{ where $\epsilon' = \epsilon e_{L}/(2L)$} \end{equation}
and $L = \sup_{0 \leq y \leq 1} l(y)$.

Suppose the $r$th word in the Lempel-Ziv parsing begins at $t(r)$, has
length $s(r)$ and ensemble-entropy defined to be
$ \ent^{(\ell)}(r) = \sum_{u=t(r)}^{t(r)+s(r)-1}  g(u/\ell).$
We set $N = N (\vec{Y}^{(\ell)}) = \{ t(r): \ent^{(\ell)}(r)
\leq \log \ell(1- \epsilon) \}$ (the set of
start-points of low ensemble-entropy words).
For any sequence ${\vec{k}}$
we can write:
\begin{eqnarray}
N & \subseteq &
\left\{ t(r) : \ent^{(\ell)}(r) \leq \log \ell(1-\epsilon),
\vec{k}\in {\mathcal{T}}^{(\ell)}_{t(r),s(r)} \right\} \nonumber \\
& &  
\bigcup  \left\{ t(r): \vec{k} \notin {\mathcal{T}}^{(\ell)}_{t(r),s(r)} 
\right\} \label{eq:nsplit}.
\end{eqnarray}
We bound the size of the first set in Proposition \ref{prop:intyp},
to find that it is less than $K_1 \ell^{1-\epsilon^2}$,
since these parsed words are short distinct strings in the typical set. 
\begin{proposition} \label{prop:intyp}
Given $\epsilon$, 
if $l(y)$ is uniformly continuous on $[0,1]$ and 
bounded above by $L$
there exists a constant $K_1(\epsilon)$ such that:
\begin{equation}
\#\left\{r : \ent^{(\ell)}(r) \leq \log \ell(1-\epsilon),
\vec{k}\in {\mathcal{T}}^{(\ell)}_{t(r),s(r)} \right\} \leq K_1 
\ell^{1-\epsilon^2}.\end{equation}
\end{proposition}
\begin{proof}
We can find a finite number of intervals $J_i$ in 
which our variables have 
their means close together. The key property is 
that $l(y)$ is (uniformly) continuous, so 
given $\epsilon$, we can calculate 
$N=N(\epsilon)$ and $u_1,\ldots u_N$ with $u_1 = 0$, $u_N= 1$ 
such that for $i=1,\ldots N-1$:
$$\sup_{x,y \in [u_i,u_{i+1}]}\left|l(x) - l(y)\right| \leq \epsilon'.$$
where $\epsilon'$ is from Equation (\ref{eq:typset}).
Define $J_i = \{ j: j/\ell \in (u_i, u_{i+1}) \}$, 
$L_i = \sup_{x \in J_i} l(x)$.

As in \cite{johnson10},
we compare ${\mathcal{T}}^{(\ell)}_{j,M}$ with
${\mathcal{D}}_{a,M}$, a set which we can count and control more easily.
Given $a >0$ and integer $M$, define:
\begin{equation}
{\mathcal{D}}_{a,M} = \left\{ \vec{x}_1^M = (x_1,\ldots ,x_M) \in \{ 0,1 \}^M:
\sum_{i=1}^M x_i \leq M \left( a + \frac{\epsilon e_a}{a} \right) 
\right\}.\end{equation}
For $\vec{x}_1^M \in {\mathcal{D}}_{a,M}$, writing $\pr_a$ for
product measure for independent $0,1$-valued random variables with
mean $a$:
$$ \pr_{a}(\vec{x}) = \exp
\left(  M \log (1- a) + \log a \sum_{i=1}^M x_i \right) \geq 
\exp(-M e_a (1 + \epsilon)).$$ 
If $\vec{k}\in {\mathcal{T}}^{(\ell)}_{j,M}$, where $j \in J_i$:
\begin{eqnarray*} \sum_{u=j}^{j+M-1} k_u & \leq &
\sum_{u=j}^{j+M-1} l(u/\ell) + M \epsilon' \\  
& \leq & M \left( L_i  + 2\epsilon' \right)
\leq M \left( L_i + \epsilon e_{L_i}/L_i \right),\end{eqnarray*}
so the sub-string
$\vec{k}_j^{j+M-1} = (k_j, \ldots k_{j+M-1}) \in {\mathcal{D}}_{L_i,M}$.
We therefore know 
that if $\vec{k}\in {\mathcal{T}}^{(\ell)}_{t(j),t(j)+s(j)-1}$  and $s(j)
\leq (\log \ell )(1-\epsilon)/e_{j}^{(\ell)} = M(j)$ then 
\begin{eqnarray*} 
\pr_{L_i} \left( k_{t(j)},\ldots , k_{t(j)+l(j)-1} \right) & \geq & 
\pr_{L_i}\left( k_{t(j)},\ldots ,k_{t(j)+M(j)-1} \right) \\
& \geq & \exp(-M(j) e_{L_i} (1+\epsilon))
= \ell^{1 - \epsilon^{2}}.\end{eqnarray*}
Since these finite strings are distinct, the 
number of strings in $J_i$ such
that these two conditions hold is less than $\ell^{1-\epsilon^{2}}$. 
Summing over
intervals $J_i$, the total number of such strings is less than
$\ell^{1-\epsilon^{2}} N$. 
\end{proof}
\Section{Negative association and the typical set} \label{sec:na}
In this section, we develop necessary technical tools to work with the
canonical  Gibbs distribution,  and  then finish  the  proof of  Lemma
\ref{lem:firstbd}.  The  key property  that we shall  use is  that our
variables  $\vc{K}$  are negatively  associated.  That  is, under  the
condition that $\sum_{i=0}^{\ell -1}  K_i = n_{\ell}$, since $K_i$ are
non-negative, if one variable is  large, then the others are forced to
be smaller. Formally:
\begin{definition} \label{def:negass}
A collection of real-valued 
random variables $(U_k)$ is negatively associated (NA) if the covariance 
$$ \cov( f(U_i: i \in A), g(U_j: j \in B)) \leq 0,$$
for all increasing functions $f$ and $g$, taking arguments over disjoint sets
of indices $A$ and $B$. \end{definition}
We require a result that gives a class of variables with conditional 
distributions that are negatively associated. This comes via the idea of
logarithmic concavity:
\begin{definition} \label{def:logconc}
A random variable $V$ taking values in $\Z_+$
with probabilities $p(s) = \pr(V=s)$ satisfies logarithmic 
concavity (LC) if for all $s \geq 1$, $p(s)^2 \geq p(s-1) p(s+1)$.
\end{definition}
Notice that the $0,1$-valued and geometric distributions have this property.
This is sometimes referred to as Newton's inequality
(see Niculescu \cite{niculescu}).
Further, we can use the following fact, going back to Hoggar (see
\cite{hoggar}).
\begin{theorem} If $V$ and $W$ are independent LC random variables, 
then their sum $V+W$ is also LC.
\end{theorem}
We also rely on a result of Joag-Dev and Proschan \cite{joag-dev},
which is itself based on Efron \cite{efron}. We reproduce its proof
here since the proofs in \cite{efron} and
\cite{joag-dev} only describe the case of
random variables with densities.

\begin{proposition} \label{prop:negass}
Let $V_i$, $i = 1,2 \ldots$  be independent $\Z_+$-valued
LC random variables, with sum $S_{\ell} = \sum_{i=1}^\ell V_i$.
Then for any $\ell$ and $n$, the conditional variables
$W_i \sim (V_i | S_{\ell} =n)$, $i = 1, \ldots \ell$, form an NA family.
\end{proposition}
\begin{proof}
First, we establish an assertion similar to the main theorem of \cite{efron}: 
if $V_1, \ldots V_{\ell}$ are LC random variables then
for any increasing function $\phi$ the conditional expectation
\begin{equation} \label{eq:incphi}
 \ep \left( \phi(V_1, \ldots V_\ell) | S_\ell =s \right) \mbox{ is an increasing
function of $s$.} \end{equation}
We prove (\ref{eq:incphi}) by induction on $\ell$.

By log-concavity, $p_2(s+1-x)/p_2(s+1-y) \leq p_2(s-x)/p_2(s-y)$ for integer
$0 \leq x \leq y \leq s$. Then for any $1 \leq t \leq x-1$:
\begin{eqnarray*}
\frac{ \sum_{x=0}^t p_1(x) p_2(s+1-x)}{ 
 \sum_{x=t}^{s+1} p_1(x) p_2(s+1-x)}
& = & \frac{ \sum_{x=0}^t p_1(x) (p_2(s+1-x)/p_2(s+1-t)) }
{ \sum_{x=t}^{s+1} p_1(x) (p_2(s+1-x)/p_2(s+1-t))} \\
& \leq & \frac{ \sum_{x=0}^t p_1(x) (p_2(s-x)/p_2(s-t)) }
{ \sum_{x=t}^{s} p_1(x) (p_2(s-x)/p_2(s-t))} \\
& = & \frac{ \sum_{x=0}^t p_1(x) p_2(s-x)}{ 
 \sum_{x=t}^{s} p_1(x) p_2(s-x)}.
\end{eqnarray*}
Now, since $a/b \leq c/d$ implies that $a/(a+b) \leq c/(c+d)$, this gives us
that for any $t$, $\pr(V_1 \leq t | V_1 + V_2 = s+1)
\leq \pr(V_1 \leq t | V_1 + V_2 = s)$, which implies Equation (\ref{eq:incphi})
for $\ell =2$.

For $\ell > 2$ and an increasing function $\phi$ define
$$ \Phi(t,u) = \ep \left( \phi(V_1, \ldots V_\ell) | T =t, V_\ell =u 
\right), \mbox{where $T = \sum_{i=1}^{\ell-1} V_i$.}$$ 
We know that $\Phi$ is increasing in $t$
by the inductive hypothesis for Equation (\ref{eq:incphi}) for $\ell-1$, 
and in $u$ by the monotonicity of $\phi$.
Then
$$ \ep \left( \phi(V_1, \ldots V_\ell) | S_{\ell} =s \right) = 
\ep (\Phi(T,V_\ell) | T+V_\ell = s),$$
which is increasing in $s$ by the inductive hypothesis for $\ell=2$. This
concludes the proof of (\ref{eq:incphi}).

Next, as in \cite{joag-dev}, we use (\ref{eq:incphi}), relying on two further 
results.
Firstly, Chebyshev's rearrangement Lemma: for $F_+$ increasing
and $F_-$ decreasing, 
\begin{equation} \cov(F_+(X), F_-(X)) \leq 0 \label{eq:chebre} \end{equation}
(as $\ep F_+(X) \ep F_-(X) - \ep F_+(X) F_-(X) 
= \sum_{i \neq j} \left( p(i) p(j) F_+(i) F_-(j) \right.$ \\ $
\left. - p(i) p(j) F_+(i) F_-(i) 
\right)
= \sum_{j < i} p(i) p(j) (F_+(i) - F_+(j))(F_-(j)- F_-(i)) \geq 0$, where
$i,j \in \Z_+$).

Secondly, by expanding with conditioning, for any random variables $U,V,W$:
$$ \cov( U,V) = \ep \cov(U,V | W) + \cov( \ep(U|W), \ep(V|W)).$$
Now, taking $U = f(V_i, i \in A) | S_{\ell}$, $V = g(V_j, j \in B) | S_{\ell}$ 
and
$W = (S_A,S_B) = (\sum_{i \in A} V_i, \sum_{j \in B} V_j)$
where $A,B \subset \{ 1, \ldots \ell \}$ are disjoint sets:
\begin{eqnarray*} 
 \cov( f, g | S)
& = & \ep (\cov( f,g) | S, S_A, S_B) + \cov( \ep(f | S_A,S_B), 
\ep (g | S_A, S_B) |S ).
\end{eqnarray*}
The first term is zero. As for
the second term; as $S_A$ increases, \ref{eq:incphi}) implies that
$\ep(f | S_A)$ increases. At the same time, since $S_A + S_B = S_{\ell}$, $S_B$
decreases, so again by (\ref{eq:incphi}), $\ep (g | S_B)$ decreases, so we 
can apply (\ref{eq:chebre}). This completes the proof of Proposition
\ref{prop:negass}
\end{proof}
In Proposition \ref{prop:outtyp} we will work with a general family of 
$\Z_+$-valued variables, but assume that functions $l$ and $g$ satisfy
Assumption \ref{ass:mu}. 
\begin{proposition} \label{prop:outtyp}
Consider a triangular array of $\Z_+$-valued random variables 
$Y_j^{(\ell)}$, $j \in \Lambda_{\ell} = \{ 0, \ldots \ell -1 \}$,
with $Y_{\cdot}^{(\ell)}$ forming an NA
family for each $\ell$. Assume $Y_j^{(\ell)}$ have
mean $l(j/\ell)$ and entropy $g(j/\ell)$
 and a uniform bound on their centred fourth moment:
$ \ep (Y_j^{(\ell)} - l(j/\ell))^4 \leq b$.
Then for any $\epsilon$ and any $\eta \in (0,1)$, there exists a constant
$K_2 = K_2(\epsilon, \eta, b)$ such that for any $\ell$ large and
for $C$ the number of words in the Lempel-Ziv parsing:
\begin{equation}
 \pr \left( \frac{1}{C} \sum_{r=1}^C 
I(\vec{Y} \notin {\mathcal{T}}^{(\ell)}_{t(r),s(r)}) \geq \eta \right) 
\leq \frac{K_2}{C^2}.\end{equation}
Here ${\mathcal{T}}^{(\ell)}_{i,M} \subseteq \Z_+^{\ell}$ is defined in
(\ref{eq:typset}).
\end{proposition}
\begin{proof}
By NA, for any $i \neq j \neq k \neq m$ from $\Lambda_{\ell}$:
\begin{eqnarray*}
0 & \geq &
\ep (Y_i^{(\ell)} - l(i/\ell)) (Y_j^{(\ell)} - l(j/\ell))^3, \\
0 & \geq & \ep (Y_i^{(\ell)} - l(i/\ell)) (Y_j^{(\ell)} - l(j/\ell)) 
(Y_k^{(\ell)} - l(k/\ell))^2, \\
0 & \geq & \ep (Y_i^{(\ell)} - l(i/\ell)) (Y_j^{(\ell)} - l(j/\ell))
(Y_k^{(\ell)} - l(k/\ell))(Y_m^{(\ell)} - l(m/\ell)).
\end{eqnarray*}
Hence, for any set $A \subseteq \Lambda_{\ell}$:
\begin{eqnarray*}
\lefteqn{ \ep \left( \sum_{j \in A} \left(Y_j^{(\ell)} - l(j/\ell)
\right) \right)^4} \\ 
& \leq & \sum_{j \in A} \left( Y_j^{(\ell)} - l(j/\ell) \right)^4  + 3
\sum_{i,j \in A, i \neq j} \left(Y_i^{(\ell)} - l(i/\ell) \right)^2 
\left(Y_j^{(\ell)} - l(j/\ell) \right)^2 \\
& \leq & 3b | A |^2.\end{eqnarray*}
Define $Z_r = I( \vec{Y}^{(\ell)} 
\notin {\mathcal{T}}^{(\ell)}_{t(r),s(r)} )$. Note that by 
Chebyshev's inequality, for any $s$:
\begin{eqnarray*}
 \ep Z_r = \pr( \vec{Y}^{(\ell)} 
\notin {\mathcal{T}}^{(\ell)}_{t(r),s} )
& \leq &  \frac{ \ep
\left( \sum_{j=t(r)}^{t(r)+s-1} (Y_j^{(\ell)} - l(j/\ell))) \right)^4}
{(s \epsilon)^4} \leq  \frac{3b }{s^2 \epsilon^{'4}}. 
\end{eqnarray*}
Hence, it is sufficient to show that $\sum_{r=1}^C s(r)^{-2}/C \rightarrow 0$
almost surely, so that
\begin{equation}
\mbox{if $E_C = \sum_{r=1}^C \ep Z_r$, then } 
E_C/C \rightarrow 0. \label{eq:meandom}
\end{equation}
However, the counting argument described in Chapter II of \cite{shields3} 
shows that this will hold. Specifically, in the finite alphabet case, the 
number of possible parsed words of length less than $(\log \ell)/100$ is
$c \log \ell \ell^{1/100}$. In the infinite alphabet case, we can truncate 
as in \cite{johnson10}.

Then, by NA, writing $\ov{Z}_j$ for a random variable with the same marginal
distribution as $Z_j$, but independent of the other $\ov{Z}_k$:
\begin{eqnarray*}
\pr \left( \sum_{r=1}^C (Z_r - \ep Z_r)  \geq \nu C \right)
& \leq & \pr \left( \sum_{r=1}^C (\ov{Z}_r - \ep Z_r) \geq \nu C \right)
\leq  \exp \left( - 2 C \nu^2 \right) 
\end{eqnarray*}
by Hoeffding's inequality (see for example Note 2.6.2 of \cite{petrov}).

Hence, writing $\eta C = \nu C + E_C$
$$ \pr \left( \sum_{r=1}^C Z_r \geq \eta C \right)
\leq  \exp \left( - 2 C (\eta - E_C/C)^2 \right) \leq  
\frac{1}{2 C^2 (\eta - E_C/C)^4},$$
so by (\ref{eq:meandom}), we are done.
\end{proof}
We can now complete the proof of Lemma \ref{lem:firstbd}:

\begin{proof}{\bf of Lemma \ref{lem:firstbd}}
We need to bound the $|N|$, by controlling the 
two sets in the RHS of (\ref{eq:nsplit}). Proposition
\ref{prop:intyp} gives that the size of the first set is 
$O(\ell^{1 - \epsilon^2})$.

Proposition \ref{prop:negass} shows that our variables have the NA property
and hence we can apply Proposition \ref{prop:outtyp}
to the size of the second set in the RHS of (\ref{eq:nsplit}).
Specifically, if 
$C(\vec{Y}^{(\ell)})$ grows more slowly than 
linearly in $\ell/\log \ell$, then Theorem \ref{thm:main} holds. Otherwise, 
the upper bound $O(1/C^2)$ provided by
of Proposition \ref{prop:outtyp} becomes 
summable in $\ell$. Hence by the Borel-Cantelli Lemma, 
almost surely the proportion
of words in the second set becomes smaller than $\eta$, for any $\eta$.
Overall then, the $|N|\log \ell/\ell \rightarrow 0$.

Then considering the entropy present in the parsed words, we deduce that:
$$ \sum_{j=1}^{\ell} g(j/\ell) = \sum_{r} 
\ent^{(\ell)}(r)
\geq \log \ell (1-\epsilon) (C(\vec{Y}^{(\ell)}) - |N|). $$
On rearranging we obtain that
$$\limsup_{\ell\to\infty} \frac{\log \ell}{\ell} 
C(\vec{Y}^{(\ell)})
\leq \frac{1}{\ell} \sum_{j} g(j/\ell) +\epsilon.$$ 
\end{proof}

\Section{The von Neumann entropy of the ensemble} \label{sec:vnaens}
One might expect that the conditioning might have only a small effect on the 
entropy of the ensemble, and that hence the entropy of the canonical and 
grand canonical ensembles will be very close to one another. This is 
confirmed in this section, in which we prove Lemma \ref{lem:secondbd}.

In the IID case of $0,1$-valued random variables, 
it is clear that Assumption \ref{ass:nogrow} is the right
condition, since there the conditioned variables
$ Y_j^{(\ell)} \sim (K_j^{(\ell)} | \sum_{0 \leq i \leq \ell-1} K_i = 
n_{\ell})$ is equiprobable
on the $\bin{\ell}{n_\ell}$ possible values, so
\begin{eqnarray*}
\lefteqn{H \left( K_0, \ldots, K_{\ell-1} \left| \sum_{0 \leq i \leq \ell-1} K_i = 
n_\ell \right) \right. - H(K_0, \ldots, K_{\ell-1}) } \\
& = & \log \bin{\ell}{n_\ell} - \ell (-p \log p - (1-p) \log (1-p)) \\
& \simeq & \left( \frac{\ell}{2p(1-p)} \right) \left( \frac{n_\ell}{\ell} -p
\right)^2 \rightarrow 0 \mbox{ as $\ell \tends$.}
\end{eqnarray*}
 (using Stirling's formula, and 
expanding in a series in $n_\ell/\ell$ close to $p$).

In the non-IID case
we exploit a relationship between the mode and mean described by Bottomley
\cite{bottomley}. Specifically, for a unimodally distributed 
random variable $S$:
$$ |\rm{mode}(S) - \ep S| \leq \sqrt{3 \var S}.$$
Hence if $S$ is the sum of $n$ `approximately IID' variables, we expect that
the mean will be of the order of $n$, and within $\sqrt{n}$ of the mode.

As elsewhere in this paper (and in \cite{johnson10}), we will use the fact that
along small enough intervals the variables are `nearly IID'. That is, by
continuity of the mean-value function $l$, we can find intervals such that 
$\sup_{y,y' \in I_j} |l(y) - l(y')|$ is arbitrarily close to zero. 
\begin{lemma} \label{lem:nearlycon}
Given $\epsilon >0$,
we can find a partition of $[0,1]$ by intervals $I_j = [u_j,u_{j+1}]$ such that defining 
$$ k^*_{j,\ell} = \frac{1}{(u_{j+1} - u_j)\ell} \sum_{i/\ell \in I_j} 
\frac{l(i/\ell)}{1-l(i/\ell)},$$
then $$l_j \leq 
\lim_{\ell \rightarrow \infty} k^*_{j,\ell}/(1+ k^*_{j,\ell}) \leq l_j (1+\epsilon).$$
where $l_j = \int_{I_j} l(x) dx$ 
\end{lemma}
In the case of $0,1$-variables, we require an extra statement, as follows
(the case of geometric variables is actually simpler, and discussed at the end
of the section).
\begin{lemma} \label{lem:scorebd}
For the sum $S^{(M)}$ of independent $0,1$-variables $X_1, \ldots X_M$, with $\ep
X_j = p_j$,
define $k^*_M = \left( \sum_{i=1}^M p_i/(1-p_i) \right)/M$, then:
$$ \frac{ \pr( S^{(M)} = n-1)}{\pr(S^{(M)} =n)} \geq \frac{n}{(M-n+1) k^*_M}.$$
\end{lemma}
\begin{proof}
Following Niculescu \cite{niculescu}, write $E_i$ for the elementary symmetric
functions of degree $i$, and ${\cal E}_i = E_i/\bin{M}{i}$ for the averaged version.
By Newton's inequalities, discussed in \cite{niculescu}, 
the ratio ${\cal E}_{n-1}/{\cal E}_n
\geq {\cal E}_0/{\cal E}_1$. Hence, we deduce that, writing $u_j = p_j/(1-p_j)$:
\begin{eqnarray*}
\frac{ \pr( S^{(M)} = n-1)}{\pr(S^{(M)} =n)} & = & 
\frac{E_{n-1}(u_1, \ldots u_M)}{E_{n}(u_1, \ldots u_M)} =
\frac{\bin{M}{n-1} {\cal E}_{n-1}(u_1, \ldots u_M)}{\bin{M}{n} 
{\cal E}_{n}(u_1, \ldots u_M)} \\
& = & \frac{n {\cal E}_{n-1}(u_1, \ldots u_M)}{(M-n+1) {\cal E}_{n}(u_1, \ldots u_M)} \\
& \geq & \frac{n {\cal E}_{0}(u_1, \ldots u_M)}{(M-n+1) {\cal E}_{1}(u_1, 
\ldots u_M)} = \frac{n}{(M-n+1) k^*_M}.
\end{eqnarray*}
\end{proof}
We will now use a local lattice Central Limit Theorem, the main result
of Petrov \cite{petrov2} (see Theorem \ref{thm:petrov} below).
Note that this result extends a very similar one of Prohorov \cite{prohorov2},
which only holds for uniformly bounded variables, thus ruling out the
geometric case.
\begin{assumption} \label{ass:gdbeh} Consider integer-valued
random variables $X_1, X_2, \ldots$ satisfying the following conditions:
\begin{enumerate}
\item{The highest common factor of the integers $j$ such that
$$ \frac{1}{\log n} \left( \sum_{i=1}^n \pr(X_i = 0) \pr(X_i = j) \right)
\rightarrow \infty$$ is 1, where for each $X_i$, by shifting, 
$\pr(X_i = j)$ has its largest value at $j=0$.}
\item{As $n \tends$, $\sigma^2_n \tends$, and $\sup_m L_m \sigma_m < \infty$
where $\sigma^2_n = \sum_{i=1}^n \var X_i$ and
\begin{equation} \label{eq:gdbeh}
L_n = \frac{1}{\sigma_n^3} \sum_{i=1}^n \ep | X_i - \ep X_i |^3.
\end{equation}}
\end{enumerate}
\end{assumption}

\begin{theorem}[Petrov] \label{thm:petrov}
Let $X_1, X_2, \ldots$ be
independent integer-valued random variables 
and write $S^{(n)} = \sum_{i=1}^n X_i$, $a_n = \ep S^{(n)}$ and $\sigma^2_n
= \var S^{(n)}$. Then
under Assumption \ref{ass:gdbeh} there exists a constant $C$ such that
for all $q in \Z$:
$$ \left| \sigma_n \pr(S_n = q) - \frac{1}{\sqrt{2 \pi}} \exp
\left( - \frac{ (q-a_n)^2}{2\sigma^2_n} \right) \right| \leq C L_n,$$
where $L_n$ is defined in (\ref{eq:gdbeh}) above.
\end{theorem}

To apply Theorem \ref{thm:petrov} we need to
check that Assumption \ref{ass:gdbeh}.2 holds in both the
fermionic and bosonic case.
For a $0,1$-variable $X$ with $\ep X = p$, $\ep|X - p|^3 = 
p(1-p)(2p^2 - 2p +1) \leq 2 \var X$. Hence in the fermion case, variables
$K_0, \ldots K_{\ell -1}$ satisfy:
\begin{equation} \label{eq:k3bd} 
\sum_{0 \leq i \leq \ell -1} \ep | K_i - \ep K_i |^3
\leq c \sum_{0 \leq i \leq \ell -1} \var K_i \end{equation}
with $c=2$ and Assumption \ref{ass:gdbeh}.2 holds.

For $X$ geometric, we can distinguish two cases. (a) Parameter
$p < 1/2$, so that
$\ep X = p/(1-p) < 1$ and 
$\ep |X-\ep X|^3 = \ep (X - \ep X)^3 + 2 \ep X^3 \pr(X =0)
= p(1+p+2p^2 -2p^3)/(1-p)^3 \leq 7 \var X$.
(b) Value $p \geq 1/2$, so $\ep X \geq 1$ and
$ \ep | X - \ep X|^3 \leq \ep X^3 + \ep X^3 = (p + 4p^2 + 2p^3)(1-p)^3
\leq 7/(1-p)^3 \leq 28 \ep X \var X$. In
either case the bound (\ref{eq:k3bd}) is fulfilled with $c = 28 L =
28 \max_i \ep K_i$ and again Assumption \ref{ass:gdbeh}.2 holds.

\begin{proof}{\bf of Lemma \ref{lem:secondbd}}
Again, let us first consider the case of $0,1$-valued variables. 
We again break the interval $[0,1]$ 
down into a collection of smaller ones as described
in Lemma \ref{lem:nearlycon}, such that the
variables are approximately IID on each interval. Refer to the sum over 
interval $I_j$ as $U_j$, and set $p_j = l(j/\ell)$.
Further, for technical reasons, we insist that the
point $1/2$ occurs at the boundary of an interval (so no interval contains
points where $p_j > 1/2$ and $p_j < 1/2$).

Writing $S^{(\ell)}$ for $\sum_{0 \leq i \leq \ell -1} K_i$, we
formally expand $H( K_0, \ldots, K_{\ell-1} 
| S^{(\ell)}=n_\ell) - H(K_0, \ldots, K_{\ell-1})$ as
\begin{eqnarray}
\lefteqn{ 
\sum_{k_0 + \ldots k_{\ell-1} = n_\ell} \frac{ \pr(K_0 = k_0) \ldots 
\pr(K_{\ell-1} = k_{\ell-1})}{\pr(S^{(\ell)} = n_\ell)} 
\left( \log \pr(S^{(\ell)} = n_\ell) 
- \sum_{0 \leq i \leq \ell-1} \log \pr(K_i = k_i) \right)} \nonumber \\ 
& & + \sum_{0 \leq i \leq \ell-1} 
\sum_{k_i} \pr(K_i = k_i) \log \pr(K_i = k_i) \nonumber \\
& = & \sum_{0 \leq i \leq \ell-1} \sum_{k_i}
\left( \pr(K_i = k_i) - \frac{\pr(K_i = k_i) \pr(S^{(\ell)}_{i} 
= n_\ell-k_i)}
{\pr(S^{(\ell)}=n_\ell)} \right) \log \pr(K_i = k_i) \label{eq:firstterm} 
\hspace*{1cm}\\
& & + \log \pr(S^{(\ell)} = n_\ell), \label{eq:secondterm}
\end{eqnarray}
where $S^{(\ell)}_{i} = \sum_{j \neq i} K_j = S^{(\ell)} - K_i$.

We deal with each of (\ref{eq:firstterm}) and (\ref{eq:secondterm})
separately. The second term,
(\ref{eq:secondterm}), is bounded directly using Theorem 
\ref{thm:petrov}, since $ (n_\ell - r)^2/\left( \sum_{0 \leq i \leq \ell-1}
\var K_i \right) \rightarrow 0$ as $\ell \rightarrow \infty$ 
(see (\ref{eq:partden2})).

Next, we split the sum over $i$ 
up into subintervals: given a value $i$, where $i/\ell
\in I_j$, we define $$U_j^{(\ell)} = \sum_{r: r/\ell \in I_j} K_r, 
\ov{U}_{j,i}^{(\ell)} 
= \sum_{r: r/\ell \in I_j, r \neq i} K_r = U_{j}^{(\ell)} - K_i, 
\overline{S}^{(\ell)}_j = 
\sum_{r: r/\ell \notin I_j} K_r = S^{(\ell)} - U_j^{(\ell)}.$$
Then the first term, Equation (\ref{eq:firstterm}) can be
rearranged to give the sum over $i$ of:
\begin{eqnarray}
\lefteqn{ \sum_t \frac{ \pr(U_{j}^{(\ell)} = t) \pr(\ov{S}^{(\ell)}_j = n-t)}
{\pr(S^{(\ell)} = n_\ell)}
\sum_k 
\left( 1 - \frac{\pr(\ov{U}_{j,i}^{(\ell)} = t-k)}{\pr(U_{j}^{(\ell)} = t)} 
\right)} \hspace*{6cm}  \nonumber \\ 
& & \times \pr(K_i= k)
\log \pr(K_i = k).\label{eq:thirdterm} \end{eqnarray}
Now, for any value of $t$, this inner sum can be rearranged to give:
\begin{equation} \label{eq:ugly}
 \log (1/p_i -1)p_i(1-p_i) 
\left( \frac{ \pr(\ov{U}_{j,i}^{(\ell)} = t-1)/\pr(\ov{U}_{j,i}^{(\ell)}
 = t) - 1}
{1-p_i + p_i \pr(\ov{U}_{j,i}^{(\ell)} = t-1)/\pr(\ov{U}_{j,i}^{(\ell)} 
= t)} \right).\end{equation}
Notice that if $t$ is close to the mode of $\ov{U}_{j,i}^{(\ell)}$, 
then this is close to zero. Hence, if the mean and mode are `close together' 
(as \cite{bottomley} ensures), we can produce sensible bounds, using
Lemma \ref{lem:scorebd}.

If $p_i \leq 1/2$, $\log(1/p_i - 1)$ is positive, and so $0 \leq (1-p_i)
p_i \log(1/p_i -1) \leq 1$.
Writing $N$ for the number of summands in $U_{j}^{(\ell)}$, Lemma 
\ref{lem:scorebd}
implies that $\pr(\ov{U}_{j,i}^{(\ell)} = t-1)/\pr(\ov{U}_{j,i}^{(\ell)} 
= t) \geq t/(k^*(N-t))$, so since
$f(v) = (v-1)/(1-p+p v)$ is an increasing function in $v$, we can deduce
that (\ref{eq:ugly}) is at least:
\begin{eqnarray*}
\frac{ 
t(k^* + 1) - k^* N}{k^* N (1-p) - t (k(1-p) - p)} 
& \geq & \frac{(1+k^*)^2}{k^* N} \left( t - \frac{k^* N}{k^* + 1} \right) 
\geq  \frac{4}{N} \left( t - \frac{k^* N}{k^* + 1} \right), \nonumber  
\end{eqnarray*}
where the first inequality follows by concavity in $t$.

Thus we control (\ref{eq:firstterm}) through a bound on
$ \ep \left( U_{j}^{(\ell)} - \ep U_{j}^{(\ell)} 
 \Big| S^{(\ell)} = n_\ell\right)$
provided by Theorem \ref{thm:petrov}.
That is, we know that $U_{j}^{(\ell)}$, $\overline{S}^{(\ell)}_j$ 
and $S^{(\ell)}$ are
all close to normal, so the expectations will be close to the values they
take in the normal case. That is, writing $Z_1, Z_2$ and $Z_3$ for normal
variables, with densities
$\phi_1$, $\phi_2$ and $\phi_3$ of mean and variance matching
$U_{j}^{(\ell)}, \overline{S}^{(\ell)}_j$ and $S$ respectively:
\begin{eqnarray*}
\lefteqn{ \left| \sum_y \frac{ y \pr( U_{j}^{(\ell)} = y) 
\pr(\overline{S}_j =n_\ell-y)}{\pr(S^{(\ell)} =n_\ell)}
- \frac{ y \phi_1(y) \phi_{2}(n_\ell-y)}{\phi_3(n_\ell)} \right|} \\
& \leq & \sum_y \frac{ |y \pr( U_{j}^{(\ell)} =y)| 
|\pr(\overline{S}^{(\ell)}_j = n_\ell-y) - \phi_{2}(n_\ell-y)|}
{\pr(S^{(\ell)} =n_\ell)} \\
& & +  \sum_y \frac{ |y \phi_{2}(n_\ell-y)| |\pr(U_{j}^{(\ell)} = y) - 
\phi_{1}(y)|} {\pr(S^{(\ell)} =n_\ell)}  \\
& & + \sum_y \frac{ |y| \phi_1(y) \phi_{2}(n_\ell-y)}{\pr(S^{(\ell)} = n_\ell) \phi_3(n_\ell)} 
\left| \pr(S^{(\ell)} = n_\ell) - \phi_3(n_\ell) \right| \\
& \leq & \frac{\epsilon}{\pr(S^{(\ell)} = n_\ell)} 
\left( \sqrt{ \var U_{j}^{(\ell)}} + |n_\ell- \ep \overline{S}^{(\ell)}_j| 
+ \sqrt{ \var \overline{S}^{(\ell)}_j} 
+ \sqrt{ \var U_{j}^{(\ell)}} \right) \\
& = & O \left( \frac{1}{\sqrt{\ell}} \right),
\end{eqnarray*}
where $\epsilon$ is the largest of the bounds given
by Theorem \ref{thm:petrov}, and hence is $O(1/\ell)$.

The result follows, using the fact that if $Z_i$ are independent $N(\mu_i,
\sigma^2_i)$ random variables (for $i=1,2$) 
then $Z_1 \big| Z_1 + Z_2 = \mu_1 + \mu_2 + r$
has a $N(\mu_1 + r\sigma^2_1/(\sigma^2_1 + \sigma^2_2),
\sigma^2_1 \sigma^2_2/(\sigma^2_1 + \sigma^2_2))$ distribution.

For $p_i > 1/2$, the same argument works, only replacing the lower bound on
$ \pr(n-1)/\pr(n)$ from Lemma \ref{lem:scorebd} with the corresponding upper
bound that $\pr(n-1)/\pr(n) \leq l^* n/(N-n+1)$, where 
$l^* = (\sum (1-p_i)/p_i)/N$. The argument goes through in a similar way.
\end{proof}

We can use a similar idea in the case of geometric distributions.
Again we can bound the term (\ref{eq:secondterm}) using Theorem 
\ref{thm:petrov}.
Then the first term, Equation (\ref{eq:firstterm}) can be
rearranged to give the sum over $i$ of:
\begin{equation}
 \sum_t \frac{ \pr(\ov{U}_{j,i}^{(\ell)} = t) \pr(\overline{S}_j = n-t)}{\pr(S^{(\ell)} = n)}
\sum_x x \left( \pr(X_i = x)
 - \frac{\pr(X_i = x) \pr(T'_j = t-x)}{\pr(\ov{U}_{j,i}^{(\ell)} = t)} \right) 
,\end{equation}
which is the same term previously bounded, and the same arguments apply.

\section*{Acknowledgements}
YS thanks IHES, Bures-sur-Yvette, for hospitality during visits in 2002
and 2003. 
OJ is a fellow of Christ's College Cambridge. Both 
authors are part of the Cambridge-MIT Institute collaboration `Quantum
Information Theory and Technology'.

\end{document}